\pdfoutput=1
\documentclass[10pt,conference]{IEEEtran}
\usepackage[T1]{fontenc}
\usepackage[utf8]{inputenc}
\usepackage[english]{babel}

\usepackage[style=ieee, maxnames=9, minnames=9, date=year, doi=false, isbn=false, url=false, backend=biber, sortcites]{biblatex}
\usepackage{nicefrac}
\usepackage{mathtools,amssymb,amsfonts}
\usepackage{csquotes}
\usepackage{graphicx}
\usepackage{subcaption}
\usepackage{multirow, booktabs}
\usepackage[detect-all,binary-units=true]{siunitx}
\usepackage[usenames,dvipsnames]{xcolor}
\usepackage{tikz}
\usepackage[textsize=scriptsize]{todonotes}
\usepackage{microtype}
\usepackage{algorithm,algpseudocode}
\usepackage{hyperref}

\usetikzlibrary{calc, positioning, fit}

\makeatletter
\let\oldtheequation\theequation
\renewcommand\tagform@[1]{\maketag@@@{\ignorespaces#1\unskip\@@italiccorr}}
\renewcommand\theequation{(\oldtheequation)}
\makeatother

\newtheorem{example}{Example}

\newcommand*{\ket}[1]{\ensuremath{|#1\rangle}}
\AtBeginDocument{}

\hypersetup{
	pdftitle={Adaptive Compilation of Multi-Level Quantum Operations},
	pdfsubject={IEEE International Conference on Quantum Computing and Engineering (QCE) 2022},
	pdfauthor={Kevin Mato, Martin Ringbauer, Stefan Hillmich, and Robert Wille} 
}

\begin{document}
\title{Adaptive Compilation\\ of Multi-Level Quantum Operations}
\author{
\IEEEauthorblockN{%
	Kevin Mato\IEEEauthorrefmark{1}\hspace{1cm}%
	Martin Ringbauer\IEEEauthorrefmark{2}\hspace{1cm}%
	Stefan Hillmich\IEEEauthorrefmark{3}\hspace{1cm}%
	Robert Wille\IEEEauthorrefmark{1}\IEEEauthorrefmark{4}%
	}
\IEEEauthorblockA{\IEEEauthorrefmark{1}Chair for Design Automation, Technical University of Munich, Munich, Germany}
\IEEEauthorblockA{\IEEEauthorrefmark{2}Institut für Experimentalphysik, University of Innsbruck, Innsbruck, Austria}
\IEEEauthorblockA{\IEEEauthorrefmark{3}Institute for Integrated Circuits, Johannes Kepler University Linz, Linz, Austria}
\IEEEauthorblockA{\IEEEauthorrefmark{4}Software Competence Center Hagenberg (SCCH) GmbH, Hagenberg, Austria}
\IEEEauthorblockA{\href{mailto:kevin.mato@tum.de}{kevin.mato@tum.de}, \href{mailto:martin.ringbauer@uibk.ac.at}{martin.ringbauer@uibk.ac.at}, \href{mailto:stefan.hillmich@jku.at}{stefan.hillmich@jku.at}, \href{mailto:robert.wille@tum.de}{robert.wille@tum.de}}%
\IEEEauthorblockA{\url{https://www.cda.cit.tum.de/research/quantum/}}
}
\maketitle
\begin{abstract}
    Quantum computers have the potential to solve some important industrial and scientific problems with greater efficiency than classical computers. 
    While most current realizations focus on two-level qubits, the underlying physics used in most hardware is capable of extending the concepts to a multi-level logic---enabling the use of qu\emph{d}its, which promise higher computational power and lower error rates. 
    Based on a strong theoretical backing and motivated by recent physical accomplishments, this also calls for methods and tools for compiling quantum circuits to those devices. 
    To enable efficient qudit compilation, we introduce the concept of an energy coupling graph for single-qudit systems and provide an adaptive algorithm that leverages this representation for compiling arbitrary unitaries.
    This leads to significant improvements over the state-of-the-art compilation scheme and, additionally, provides an option to trade-off worst-case costs and run-time. 
    The developed compiler is available via \href{https://github.com/cda-tum/qudit-compilation}{github.com/cda-tum/qudit-compilation} under an \mbox{open-source} license.
\end{abstract}

\section{Introduction}
Quantum computers promise to solve problems of industrial and scientific interest with better resource and algorithmic efficiency. 
State-of-the-art quantum devices, subject of current research, are referred to as noisy intermediate-scale quantum (NISQ) devices~\cite{preskill2018quantum}. 
These devices host up to hundreds of noisy quantum bits (qubits) and support a limited number of logical operations on these qubits. 
Such NISQ devices have now been realized in a number of technological platforms, including superconducting circuits~\cite{Arute2019}, trapped ions~\cite{Pogorelov2021}, and single photons~\cite{Flamini2018a}. 
Notably, while these devices almost exclusively work with two-level qubits, the underlying hardware almost always natively supports encoding multiple-valued logic in \emph{qudits} (quantum digits).

The research on qudit design and computation has a long history, with efforts primarily focusing on conceptual studies of algorithms for idealized qudits and their comparison to qubits~\cite{Wang2020}. Fundamentally, a qudit can not only store and process more information per quantum particle, but also features a richer set of logical operations~\cite{Huber2013} that make information processing more efficient. As a consequence, it has been shown that qudit computation enables improvements in algorithmic and circuit complexity~\cite{Wang2020} for a wide class of problems.
These results have inspired the proposals for and demonstration of basic qudit control in numerous physical platforms, from  trapped ions~\cite{Zhang2013Contextuality,ringbauer2021universal}, to photonic systems~\cite{Lanyon2008,Ringbauer2017Coherence, Hu2018a, Malik2016}, \mbox{superconducting circuits}~\cite{Kononenko2020,Morvan2020}, Rydberg atoms~\cite{Ahn2000}, nuclear spins~\cite{Godfrin2017}, cold atoms~\cite{Anderson2015} and nuclear magnetic resonance systems~\cite{Gedik2015}. 
More recently, efforts have intensified with the demonstration of a universal qudit quantum processor with error rates that are competitive to qubit systems~\cite{ringbauer2021universal}.

One key component to utilize the full potential of qudit hardware, however, is a method to translate abstract quantum algorithms given as unitaries to the elementary logical operations of the specific hardware platform.
Due to additional structural constraints, complex cost functions, and a large number of different but valid decompositions of any given quantum operation~\cite{ringbauer2021universal,Low_2020}, compilation for qudits is significantly more complex than for binary quantum computers.
The state-of-the-art static QR decomposition~\mbox{\cite{brennen2005efficient,ringbauer2021universal,2005Brennen_Criteria,O_Leary_2006,Bullock_2005,Kiktenko_scalable,Low_2020}} fails to exploit the full potential of the underlying technology thus far (discussed and illustrated in \autoref{subsec:motivation}).

In this work, we develop an efficient adaptive compiler for single qudit operations. 
To this end, we first introduce a corresponding representation of underlying physical options and constraints (i.e., the available potential) in terms of an \emph{energy coupling graph}. 
Based on this, we afterwards propose adaptive methods that aims at leveraging this representation for efficiently compiling arbitrary unitaries. 
In particular, these methods exploit free placement of logical information on the underlying physical information carriers to achieve a significant improvement in complexity of the compiled operations. 

To showcase the benefits of the proposed methods, we compiled large sets of unitaries of dimensions 3, 5, and 7 to different target architectures---confirming the flexibility to adapt to arbitrary coupling structures of different multi-level systems as well as experimental constraints.
While the used cost-function is inspired by a recently presented qudit hardware platform~\cite{ringbauer2021universal}, the presented methods are general and apply to any physical platform and cost function.
Finally, a comparison to the state-of-the-art QR decomposition showed an significant improvement in the resulting costs on average; with further potential to trade-off worst-case costs and run-time.
The resulting tool has been made available via \href{https://github.com/cda-tum/qudit-compilation}{github.com/cda-tum/qudit-compilation} under an \mbox{open-source} license. 

\begin{figure*}[t]
    \centering
	\resizebox{0.8\linewidth}{!}{
	   \begin{tikzpicture}
            \begin{scope}[very thick]
                \draw (0,0) -- (2.5,0) coordinate[midway] (r1);
                \draw (0,1) -- (2.5,1) coordinate[midway] (r2);
                \draw (r1) -- (r2);
                \node[draw,rounded corners,thin,fill=blue,text=white] at (r1) {\(R_{0,1}(\theta, \phi)\)};
            \end{scope}
            \node[] (arrow) at (3,0.5) {\( \Longleftrightarrow \)};
            \begin{scope}[xshift=3.5cm]
                \draw[very thick] (0,0) -- (11.75,0);
                \draw[very thick] (0,1) -- (11.75,1);
                \node[thin,fill=olive,text=white] (r0a) at (1,1) {\(R_{0,\alpha}(\pi, \frac{\pi}{2})\strut\)};
                \node[thin,fill=orange,text=white, right=0.25 of r0a] (r01) {\(R_{0,1}(\theta, \phi-\frac{\pi}{2})\strut\)};
                
                \node[fill=Cerulean,text=white, right=0.25 of r01] (rms) {\phantom{\(\mathrm{MS}_{0,1}(\theta, \phi)\)}};
                \coordinate[below=of rms] (rmsb);
                \node[draw, thick, inner sep=0, fill=Cerulean, text=white, fit=(rms)(rmsb)] {};
                \node[text=white, right=0.25 of r01] (rms) {\(\mathrm{MS}_{0,1}(\theta, \phi)\)};
                \node[thin,fill=orange,text=white, right=0.25 of rms] (r01p) {\(R_{0,1}(\theta, \phi+\frac{\pi}{2})\strut\)};
                \node[thin,fill=olive,text=white, right=0.25 of r01p] (r0ap) {\(R_{0,\alpha}(\pi, \frac{\pi}{2})\strut\)};
            \end{scope}
        \end{tikzpicture}
	}
	\caption{Example of local operations on entangling operation.}
	\label{fig:entanglement}
\end{figure*}
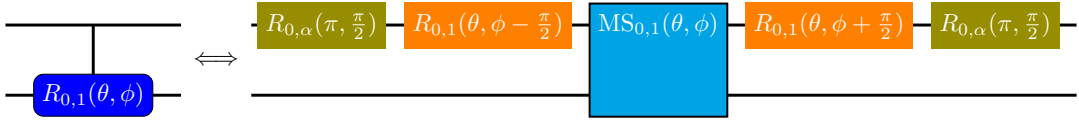

The remainder of the paper is structured as follows: \autoref{sec:background} provides the required background to keep the paper self-contained and motivates the problem.
In \autoref{sec:adaptive-compilation}, we introduce the proposed algorithm alongside the required cost~function.
\autoref{sec:evaluation} evaluates the proposed approach.
Finally, \autoref{sec:conclusion} concludes the paper.

\section{Background and Motivation}
\label{sec:background}

In this work, we provide the basis for efficient compilation of qudit operations. 
To this end, we first briefly review the basics of quantum information with a particular focus on multi-level quantum logic. 
Afterwards, we discuss the currently applied scheme of compiling corresponding operations and their drawbacks---providing the motivation of this work.

\subsection{Quantum Information Processing}
In classical computations, the primary unit of information is the bit (binary digit), which can exclusively be observed in either the 0 or the 1 state. This concept can easily be generalized to quantum computers, with the \emph{qubit} (quantum bit) as the corresponding unit of information for quantum computations. The crucial difference to the classical case, however, is that qubits can be in any linear combination of $\ket{0}$ and $\ket{1}$ (using Dirac's bra-ket notation). 

Qubits are usually constructed by restricting the natural multi-level structure of the underlying physical carriers of quantum information. These systems, therefore, natively support \mbox{multi-level logic} with the fundamental unit of information termed a \emph{qudit} (quantum digit).
A qudit is the quantum equivalent of a $d$-ary digit with $d\geq 2$, whose state can be described as a vector in the $d$-dimensional Hilbert space $\mathcal{H}_d$. 
The state of a qudit can thus be written as a linear combination $\ket{\psi} = \alpha_0 \cdot \ket{0} + \alpha_1 \cdot \ket{1} + \ldots + \alpha_{d-1} \cdot \ket{d-1}$, or simplified as vector $ \ket{\psi} = \begin{bsmallmatrix} \alpha_0 & \alpha_1 & \ldots & \alpha_{d-1}\end{bsmallmatrix}^\mathrm{T}$, where $\alpha_i \in \mathbb{C}$ are the amplitudes relative to the orthonormal basis of the Hilbert space---given by the vectors $\ket{0}, \ket{1},\ket{2},..., \ket{d-1}$.
The squared magnitude of an amplitude $|\alpha_i|^2 $ defines the probability with which the corresponding basis state $i$ will be observed when measuring the qudit. 
Since the probabilities have to add up to $1$, the amplitudes have to satisfy $\sum_{i=0}^{d-1} |\alpha_i|^2 = 1$.
\begin{example}\label{ex:state}
    Consider a system of one qudit with only three energy levels (also referred to as \emph{qutrit}).
    The quantum state $\ket{\psi} = \sqrt{\nicefrac{1}{3}}\cdot\ket{0} + \sqrt{\nicefrac{1}{3}}\cdot\ket{1} + \sqrt{\nicefrac{1}{3}}\cdot\ket{2}$ is a valid state with equal probability of measuring each basis. 
    Equivalently, the quantum state may be represented as vector $\sqrt{\nicefrac{1}{3}}\cdot \begin{bsmallmatrix} 1 & 1 & 1\end{bsmallmatrix}^\mathrm{T}$.
\end{example}

Two key properties that distinguish quantum computing from classical computing are superposition and entanglement.
A qudit is said to be in a \emph{superposition} of states in a given basis when at least two amplitudes are non-zero relative to this basis. 
\emph{Entanglement}, on the other hand, describes a form of superposition born from interactions in multi-qudit systems. Entanglement is a powerful form of quantum correlation, where the quantum information is encoded in the state of the whole system and cannot be extracted from the individual qudits anymore.
The state of a single $d$-level qudit system can be manipulated by operations which are represented in terms of \mbox{$d \times d$-dimensional} unitary matrices $U$, i.e.,~matrices that satisfy $U^\dagger U = U U^\dagger = I$.
The state after the application of $U$ can be determined by multiplying the corresponding input state from the left with the matrix~$U$.

\begin{example}
     Consider a three-level qudit (i.e.,~a qutrit) initially in the state $\ket{0}$. Applying the Hadamard operation~$H$ to it yields the output state shown before in \autoref{ex:state}, i.e.,
    \begin{align*}H\cdot \ket{0}=
        \frac{1}{\sqrt 3}
        \begin{bmatrix}
            1 & 1 & 1 \\
            1 & e^{\frac{2\pi}{3}} & e^{\frac{-2\pi}{3}} \\
            1 & e^{\frac{-2\pi}{3}} & e^{\frac{2\pi}{3}}
        \end{bmatrix} \cdot
        \begin{bmatrix}
            1 \\
            0 \\
            0 
        \end{bmatrix} 
        = \frac{1}{\sqrt 3}
        \begin{bmatrix}
            1 \\
            1 \\
            1 
        \end{bmatrix}.
    \end{align*}
\end{example}

\subsection{Compiling Multi-Level Quantum Operations}
\label{subsec:motivation}
Using the basics of quantum information processing as reviewed in the previous section, arbitrary (multi-level) quantum operations can be defined. 
However, the elementary operations available in state-of-the-art quantum computing hardware typically couple only two levels and may be subject to additional physical or practical constraints.
As a consequence, only a small set of single-qudit operations are available natively, while the vast majority require compilation into elementary operations. 
The goal of a good compiler is then to not only realize the desired computations, but also to do so in the most efficient way by fully exploiting the potential of the available technology.
Solutions designed for compiling unitaries on qubit systems are generally directed towards scalability issues and error mitigation. 
In a first step, the focus of compilation lies on single-qudit operations.
The reasons for this are three-fold: 
\begin{enumerate}
    \item One of the main advantages of qudit systems is that complicated qubit entangling operations are traded for simpler qudit local operations. A powerful local compiler is critical for exploiting this potential.
    \item Qudit systems present the possibility of performing computations between quantum systems of different dimensions and different characteristics~\cite{Lanyon2008}. This is in contrast to qubit systems where all the units are computationally identical to each other.
    \item Compilation of entangling operations relies heavily on the use of local unitaries. Hence, a local unitary compiler, such as the one we present, is a prerequisite for compiling general multi-qudit unitaries. 
\end{enumerate}

\begin{example}
    \autoref{fig:entanglement} shows an example of an entangling operation that performs a controlled arbitrary rotation; there is a total of four unitaries and one entangling gate, in this case the Mølmer–Sørensen (MS) gate~\cite{MSgate}.
    To ensure correct behavior of the MS gate as a \(\mathit{CNOT}\) gate, it requires additional local gates on each one of the qudits involved in the operation, applied just after the native operation.
    First, $d-1$ operations per qudit are required to correct intrinsic phase shifts~\cite{ringbauer2021universal}. Second, another 4 local gates turn the corrected MS gate into the desired \(\mathit{CNOT}\) gate, resulting in a total of 10 local operations for compiling one entangling gate on two ququarts (4 levels).
    Evidently, the ratio of local gates per entangling gates is rather high and not being able to compile efficiently each one of the local unitaries is an obstacle to scalability in this type of systems once full applications are deployed.
\end{example}

The state-of-the-art qudit compilation~\cite{brennen2005efficient,ringbauer2021universal,2005Brennen_Criteria,O_Leary_2006,Bullock_2005,Kiktenko_scalable,Low_2020} relies on a straight-forward \emph{QR decomposition algorithm}, which decomposes a matrix into an orthogonal matrix and an upper triangular matrix. 
In such an algorithm, the sequence of elementary two-level operations is fixed a priori and only the angles of the rotations are calculated depending on the target unitary.
Importantly, this method does not take care of particular restrictions encountered in real-world systems, for example the impossibility of directly performing certain rotations and shorter paths for routing two levels close to each other. 
An arbitrary operation will therefore in general have a better decomposition than the one generated by the fixed sequence and, hence, the blind application of more operations could lead to greater error-rates. 
Overall, the state of the art does not fully exploit the potential of current multi-level architectures---as also illustrated by the following example. 

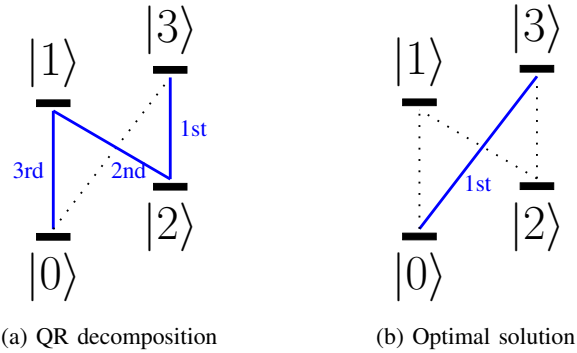
\begin{figure}[t]
	\begin{subfigure}[b]{0.45\linewidth}
	    \centering
	    \resizebox{0.7\linewidth}{!}{\begin{tikzpicture}[every label/.style={font=\huge}]
    		\draw[fill] (0.0,0.00) coordinate (r0) rectangle ++(0.5,0.1);
			\draw[fill] (1.75,0.75) coordinate (r1) rectangle ++(0.5,0.1);
			
			\draw[fill] ( 0.0,2.00) coordinate (r3) rectangle ++(0.5,0.1);
			\draw[fill] ( 1.75,2.50) coordinate (r4) rectangle ++(0.5,0.1);

			\node[thick, draw=none, inner sep=0pt, fit={($(r0)-(0.05,0.05)$)($(r0)+(0.55,0.15)$)}, label={below:$\ket{0}$}] (l0) {};
			\node[thick, draw=none, inner sep=0pt, fit={($(r1)-(0.05,0.05)$)($(r1)+(0.55,0.15)$)}, label={below:$\ket{2}$}] (l1) {};
			
			\node[thick, draw=none, inner sep=0pt, fit={($(r3)-(0.05,0.05)$)($(r3)+(0.55,0.15)$)}, label={above:$\ket{1}$}] (l3) {};
			\node[thick, draw=none, inner sep=0pt, fit={($(r4)-(0.05,0.05)$)($(r4)+(0.55,0.15)$)}, label={above:$\ket{3}$}] (l4) {};

			\draw[very thick, blue] (l0.north) -- (l3.south) node[midway, left] {3rd};
			\draw[thick, loosely dotted] (l0.north) -- (l4.south);

			\draw[very thick, blue] (l1.north) -- (l3.south) node[midway, very near start, left] {2nd};
			\draw[very thick, blue] (l1.north) -- (l4.south) node[midway, right] {1st};

	    \end{tikzpicture}}
		\caption{QR decomposition}
		\label{fig:example5-static}
	\end{subfigure}\hfill%
	\begin{subfigure}[b]{0.45\linewidth}
	    \centering
	    \resizebox{0.7\linewidth}{!}{\begin{tikzpicture}[every label/.style={font=\huge}]
    		\draw[fill] (0.0,0.00) coordinate (r0) rectangle ++(0.5,0.1);
			\draw[fill] (1.75,0.75) coordinate (r1) rectangle ++(0.5,0.1);
		    
			\draw[fill] ( 0.0,2.00) coordinate (r3) rectangle ++(0.5,0.1);
			\draw[fill] ( 1.75,2.50) coordinate (r4) rectangle ++(0.5,0.1);

			\node[thick, draw=none, inner sep=0pt, fit={($(r0)-(0.05,0.05)$)($(r0)+(0.55,0.15)$)}, label={below:$\ket{0}$}] (l0) {};
			\node[thick, draw=none, inner sep=0pt, fit={($(r1)-(0.05,0.05)$)($(r1)+(0.55,0.15)$)}, label={below:$\ket{2}$}] (l1) {};
			
			\node[thick, draw=none, inner sep=0pt, fit={($(r3)-(0.05,0.05)$)($(r3)+(0.55,0.15)$)}, label={above:$\ket{1}$}] (l3) {};
			\node[thick, draw=none, inner sep=0pt, fit={($(r4)-(0.05,0.05)$)($(r4)+(0.55,0.15)$)}, label={above:$\ket{3}$}] (l4) {};

			\draw[very thick, blue] (l0.north) -- (l4.south) node[pos=0.3, right] {1st};

			\draw[thick, loosely dotted] (l0.north) -- (l3.south) node[midway, left] {\phantom{3rd}};
			\draw[thick, loosely dotted] (l1.north) -- (l3.south);
			\draw[thick, loosely dotted] (l1.north) -- (l4.south) node[midway, right] {\phantom{1st}};

	    \end{tikzpicture}}
		\caption{Optimal solution}
		\label{fig:example5-adaptiv}
	\end{subfigure}
	\caption{Comparison between QR decomposition (3 steps) and optimal solution (1 step) to rotate between \ket0 and \ket3}
	\label{fig:example5}
\end{figure}

\begin{example}\label{ex:qr-decomposition}
    Consider a target unitary $U$ that rotates between \ket{0} and \ket{3}.
    The QR decomposition will generate three valid operations to realize the rotation, namely $U$ will be decomposed into 
    $R_{2,3},R_{1,2},R_{0,1}$.
    \autoref{fig:example5-static} illustrates the fixed sequence.
    
    Now, given the connectivity in \autoref{fig:example5}, a shorter (and optimal) decomposition $R_{0,3}$ exists, that immediately utilizes the existing connection, but is not found by the QR approach.
    \autoref{fig:example5-adaptiv} illustrates the optimal decomposition.
    In comparison, the fixed sequence requires three reordering pulses, while the optimal sequence will be composed of only one gate---highlighting the possible improvements over the state of the art even in small examples.
\end{example}

The objective of this work is to develop a method that is adaptive and can be generalized to arbitrary coupling structures. 
However, due to physical constraints, every technology has restrictions which have to be taken into account during compilation.
In order to make these restrictions transparent, we propose a dedicated representation of options as well as constraints in terms of an \emph{energy coupling graph} as well as an adaptive algorithm that enables us to control a number of \mbox{trade-offs} due to the technology that implements the qudit logic, as well as the influence of the sources of error.
Compiling in this work is then viewed as a multi-tiered procedure that requires several techniques in order to apply a single arbitrary operation as efficiently and reliably as possible. 
Moreover, an adaptive algorithm often leads to decompositions that reduce the cost of application for a given unitary, or a sequence of them. 

\section{Adaptive Compilation of\\Multi-Level Quantum Operations}
\label{sec:adaptive-compilation}
Based on the general idea sketched in \autoref{subsec:motivation}, this section provides a detailed description of the proposed approach.
To this end, we first introduce the concept of an \emph{energy coupling graph} that describes the possible physical connections in the architecture as well as the corresponding subset used for the actual compilation and, by this, provides a representation of both, the used coupling as well as the available potential.
Based on that, an adaptive decomposition approach is described which explicitly considers (is adaptive) to the potential of underlying technology when compiling a given unitary.

\subsection{Energy Coupling Graph}
\label{subsec:energy-coupling-graph}
The way information is encoded in the qudits (i.e., what energy levels represent what basis state) has a major impact on the quality of the resulting compilation result. 
Various technologies have been employed to implement quantum operations~\cite{Zhang2013Contextuality,ringbauer2021universal, Lanyon2008,Ringbauer2017Coherence, Hu2018a, Malik2016, Kononenko2020,Morvan2020, Ahn2000, Godfrin2017, Anderson2015, Gedik2015}, which leverage different physical platforms with different degrees of freedom for encoding information. 
By nature, they inherently use multi-level physical structures to represent their state, albeit currently only two of these levels are commonly used. 
The construction of a representation of these levels is compulsory for understanding how to efficiently access them, and relate to each other for performing operations.

The connections between physical levels are not arbitrary, but heavily dependent on the technology.
To capture the possible connections, we use the term \emph{coupling} for levels of a qudit in analogy to the coupling describing possible interactions between qubits in two-level quantum computers.
More precisely, a coupling between levels of a qudit is the possibility of transitioning from a certain level to a second one, given an impulse to the system. 
Furthermore, we model the possible couplings between levels with a graph---which we refer to as \emph{energy coupling graph}.
The restrictions imposed by the energy coupling graph are correspondingly referred to as \emph{energy coupling constraints}.
The term is inspired by the fact that the nodes of the graph represent different energy states within the physical information carrier.
The transitions can be mathematically represented with unitary matrix operators and associated to corresponding gates.
In terms of the two-level Bloch sphere, an operation is a rotation on the sphere with the considered energy levels at the poles. 
By that, the resulting energy coupling graph represents possible mapping of states to levels as well as correspondingly resulting constraints as provided by recent physical realizations~\cite{Low_2020}.

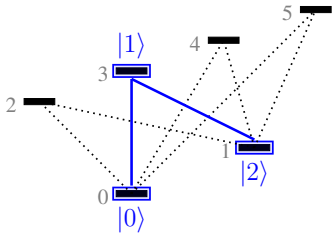
\begin{figure}[t]
    \centering
	\resizebox{0.5\linewidth}{!}{\begin{tikzpicture}
			\draw[fill] (0,0) node[left, color=gray] {0} rectangle ++(0.5,0.1) node[midway] (l0) {};
			\draw[fill] (2,0.75) node[left, color=gray] {1} rectangle ++(0.5,0.1) node[midway] (l1) {};
			\draw[fill] (0,2.0) node[left, color=gray] {3} rectangle ++(0.5,0.1) node[midway] (l3) {};
			
			\draw[fill] (-1.5,1.5) node[left, color=gray] {2} rectangle ++(0.5,0.1) node[midway] (l2) {};
			\draw[fill] (1.5,2.5) node[left, color=gray] {4} rectangle ++(0.5,0.1) node[midway] (l4) {};
			\draw[fill] (3,3.0) node[left, color=gray] {5} rectangle ++(0.5,0.1) node[midway] (l5) {};
			
			\draw[thick, draw=blue, fill=blue, fill opacity=0.1] (-0.05,-0.05) node[left, color=gray] { } rectangle ++(0.6,0.2) node[midway] (l0) {};
			\draw[thick, draw=blue, fill=blue, fill opacity=0.1] (1.95,0.70) node[left, color=gray, font=] { } rectangle ++(0.6,0.2) node[midway] (l1) {};
			\draw[thick, draw=blue, fill=blue, fill opacity=0.1] (-0.05,1.95) node[left, color=gray] { } rectangle ++(0.6,0.2) node[midway] (l3) {};
			
			\node[below=0pt of l0, color=blue, inner sep=2pt, font=\large] (k0) {$|0\rangle$};
			\node[above=0pt of l3, color=blue, inner sep=2pt, font=\large] (k1) {$|1\rangle$};
			\node[below=0pt of l1, color=blue, inner sep=2pt, font=\large] (k2) {$|2\rangle$};
			
			\draw[thick, dotted] (l0.center) -- (l2.center);
			\draw[very thick, blue] (l0) -- (l3);
			\draw[thick, dotted] (l0.center) -- (l4.center);
			\draw[thick, dotted] (l0.center) -- (l5.center);
			
			\draw[thick, dotted] (l1.center) -- (l2.center);
			\draw[very thick, blue] (l1.north) -- (l3.south);
			\draw[thick, dotted] (l1.center) -- (l4.center);
			\draw[thick, dotted] (l1.center) -- (l5.center);
	\end{tikzpicture}}
	\caption{A qutrit energy coupling graph}
	\label{fig:example3}
\end{figure}

\begin{example}
	\autoref{fig:example3} visualizes the proposed energy coupling graph for a single three-level qudit, or qutrit. 
	The black horizontal bars are the physical levels provided by the technology, and the black dotted lines show the theoretically possible couplings.
	The blue rectangles are the logical nodes of the graph mapped to a subset of the physical levels, and the blue lines are the edges of the constituted graph that are built on the theoretically available coupling.
	The blue lines represent the physically feasible transitions, or an efficient subset of them.
\end{example}

With the energy coupling graph introduced, we can graphically inspect the three main differences between the theoretically possible connectivity in a qudit and the realistic connections in a quantum system.
Given the graph of all the possible levels available from the technology, we start by mapping the logical states to the physical levels.
Typically, not all the physical levels in the energy coupling graph are used to encode logical states---some of them may be used for ancillary tasks, such as routing two distant logic states or as a cache memory, temporarily storing information.
In theory, i.e.,~in the ideal qudit, there are transitions in an all-to-all fashion between all the logic states.
However, in the non-ideal real world, only certain physical transitions can be effectively used---enabling only certain rotations in the qudit state-space. 
Additionally, there is a chance that logically contiguous states are placed on distant nodes of the graph, and it may happen when optimizing some particular applications on the qudit architecture.
In this work, we lay the foundations for a new layer of abstraction that works as an interface between a physical qudit and a corresponding compiler.

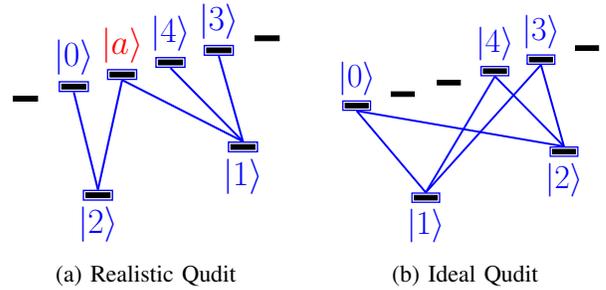
\begin{figure}[t]
    \centering
	\begin{subfigure}[b]{0.40\linewidth}
		\resizebox{\linewidth}{!}{\begin{tikzpicture}[every label/.style={blue, font=\huge}]
				\draw[fill] (0.0,0.0) coordinate (r0) rectangle ++(0.5,0.1);
				
				\draw[fill] (3.0,1.0) coordinate (r2) rectangle ++(0.5,0.1);
				
				\draw[fill] (-1.5,2.00) coordinate (r4) rectangle ++(0.5,0.1);
				\draw[fill] (-0.5,2.25) coordinate (r3) rectangle ++(0.5,0.1);
				\draw[fill] ( 0.5,2.50) coordinate (r5) rectangle ++(0.5,0.1);
				\draw[fill] ( 1.5,2.75) coordinate (r6) rectangle ++(0.5,0.1);
				\draw[fill] ( 2.5,3.00) coordinate (r7) rectangle ++(0.5,0.1);
				\draw[fill] ( 3.5,3.25) coordinate (r8) rectangle ++(0.5,0.1);

				\node[thick, draw=blue, inner sep=0pt, fit={($(r0)-(0.05,0.05)$)($(r0)+(0.55,0.15)$)}, label={below:$\ket{2}$}] (l0) {};
				\node[thick, draw=blue, inner sep=0pt, fit={($(r2)-(0.05,0.05)$)($(r2)+(0.55,0.15)$)}, label={below:$\ket{1}$}] (l2) {};
				\node[thick, draw=blue, inner sep=0pt, fit={($(r3)-(0.05,0.05)$)($(r3)+(0.55,0.15)$)}, label={above:$\ket{0}$}] (l3) {};
				\node[thick, draw=blue, inner sep=0pt, fit={($(r5)-(0.05,0.05)$)($(r5)+(0.55,0.15)$)}, label={[red]above:$\ket{a}$}] (l5) {};
				\node[thick, draw=blue, inner sep=0pt, fit={($(r6)-(0.05,0.05)$)($(r6)+(0.55,0.15)$)}, label={above:$\ket{4}$}] (l6) {};
				\node[thick, draw=blue, inner sep=0pt, fit={($(r7)-(0.05,0.05)$)($(r7)+(0.55,0.15)$)}, label={above:$\ket{3}$}] (l8) {};
				
				\draw[very thick, blue] (l0.north) -- (l3.south);
				\draw[very thick, blue] (l0.north) -- (l5.south);
				\draw[very thick, blue] (l2.north) -- (l5.south);
				\draw[very thick, blue] (l2.north) -- (l6.south);
				\draw[very thick, blue] (l2.north) -- (l8.south);
		\end{tikzpicture}}
		
		\caption{Realistic Qudit}
		\label{fig:example_ideal_vs_non-realistic}
	\end{subfigure}\qquad
	\begin{subfigure}[b]{0.40\linewidth}
		\centering
		\resizebox{\linewidth}{!}{\begin{tikzpicture}[every label/.style={blue, font=\huge}]
				\draw[fill] (0.0,0.0) coordinate (r0) rectangle ++(0.5,0.1);
				
				\draw[fill] (3.0,1.0) coordinate (r2) rectangle ++(0.5,0.1);
				
				\draw[fill] (-1.5,2.00) coordinate (r3) rectangle ++(0.5,0.1);
				\draw[fill] (-0.5,2.25) coordinate (r4) rectangle ++(0.5,0.1);
				\draw[fill] ( 0.5,2.50) coordinate (r5) rectangle ++(0.5,0.1);
				\draw[fill] ( 1.5,2.75) coordinate (r6) rectangle ++(0.5,0.1);
				\draw[fill] ( 2.5,3.00) coordinate (r7) rectangle ++(0.5,0.1);
				\draw[fill] ( 3.5,3.25) coordinate (r8) rectangle ++(0.5,0.1);
				
				\node[thick, draw=blue, inner sep=0pt, fit={($(r0)-(0.05,0.05)$)($(r0)+(0.55,0.15)$)}, label={below:$\ket{1}$}] (l0) {};
				\node[thick, draw=blue, inner sep=0pt, fit={($(r2)-(0.05,0.05)$)($(r2)+(0.55,0.15)$)}, label={below:$\ket{2}$}] (l2) {};
				\node[thick, draw=blue, inner sep=0pt, fit={($(r3)-(0.05,0.05)$)($(r3)+(0.55,0.15)$)}, label={above:$\ket{0}$}] (l3) {};
				\node[thick, draw=blue, inner sep=0pt, fit={($(r6)-(0.05,0.05)$)($(r6)+(0.55,0.15)$)}, label={above:$\ket{4}$}] (l6) {};
				\node[thick, draw=blue, inner sep=0pt, fit={($(r7)-(0.05,0.05)$)($(r7)+(0.55,0.15)$)}, label={above:$\ket{3}$}] (l8) {};
				
				\draw[very thick, blue] (l0.north) -- (l3.south);
				\draw[very thick, blue] (l0.north) -- (l6.south);
				\draw[very thick, blue] (l0.north) -- (l8.south);
				\draw[very thick, blue] (l2.north) -- (l3.south);
				\draw[very thick, blue] (l2.north) -- (l6.south);
				\draw[very thick, blue] (l2.north) -- (l8.south);
		\end{tikzpicture}}
		
		\caption{Ideal Qudit}
		\label{fig:example_ideal_vs_non-ideal}
	\end{subfigure}
	\caption{A realistic qudit, limited by physical complexity, compared to an ideal qudit, inspired by Ref.~\cite{ringbauer2021universal}}
	\label{fig:example_ideal_vs_non}
	\vspace{-0.5em}
\end{figure}

\begin{example}\label{example:example_ideal_vs_non}
	Consider \autoref{fig:example_ideal_vs_non-realistic} showing a realistic qudit, which has three main differences compared to the ideal case depicted in \autoref{fig:example_ideal_vs_non-ideal}. 
	\begin{itemize}
		\item \autoref{fig:example_ideal_vs_non-realistic} has fewer connections than the ideal case, which may be due to physical (e.g.\ atomic selection rules), or practical constraints. 
		
		\item In \autoref{fig:example_ideal_vs_non-realistic}, the logic states are not placed in order, for example the logic state $\ket2$ is connected to $\ket0$, but not $\ket1$ and $\ket3$. 
		\item Both the graphs are connected, but \autoref{fig:example_ideal_vs_non-realistic} has an ancillary state $\ket{a}$ to ensure connectivity.
	\end{itemize}
\end{example}

The energy coupling graph as proposed offers designers and automatic tools a means to properly reflect the potential as well as the restrictions of the underlying technology. 
In the following, the energy coupling graph is used to guide the compilation.

\subsection{Decomposition}

With the energy coupling graph introduced in the previous section, the next step in the multi-tiered compilation process is a \emph{decomposition} of the target unitary $U$.
In the scope of this work, the decomposition algorithm is intended to output a sequence $V_i$ of two-level rotations between adjacent physical levels in the energy coupling graph and a diagonal matrix $\Theta$ of arbitrary phases \cite{Low_2020} as in \autoref{eqn:decomposition}.
\begin{align}
    U = V_k\cdot V_{k-1}\cdot\ldots\cdot V_1\cdot\Theta\label{eqn:decomposition}
\end{align}
The objective of the algorithm is to output a sequence of operations that realize the target unitary $U$, taking into account the capabilities and restrictions of the underlying technology.

\begin{example}\label{example:example_naive_solutions}
    Consider performing a rotation between two levels, that are not directly connected.
    Some naive solutions for improving the flexibility of the QR decomposition to different graphs, could be either to permute the unitary to match the encoding on the graph, with a reordered basis set \emph{or} to fix the encoding and add single-qudit reordering pulses to the sequence to bring logic states in the energy coupling graph closer, perform the operation and then bringing them back.
    Both variants work in a \emph{static} fashion: potentially introducing unnecessary operations and, thus, unnecessary costs.
\end{example}

To alleviate the drawbacks of static decomposition approaches, we propose an \emph{adaptive algorithm} based on a recursive depth-first search and using a tree structure to keep track of the exploration of the possible decomposition. 
An overview is provided in \autoref{alg:adaptive-composition}.
By adaptively changing the energy coupling graph, this algorithm avoids the problems the static solutions in \autoref{example:example_naive_solutions} have.
The algorithm starts by allocating only the root of the tree, where the only information stored is the cost limits imposed by the fixed QR decomposition, the target unitary to compile and the initial energy coupling graph.
The cost limits for the adaptive approach are the costs of the QR decomposition applied on the same target unitary, and the summed costs of each single rotation of the adaptive decomposition will not exceed these limits.
The complexity for calculating this cost bound is asymptotically quadratic in the number of dimensions.
In every step, the algorithm applies a two-level Givens-rotation~\cite{2005Brennen_Criteria, ringbauer2021universal} to the unitary and checks if the resulting matrix is diagonal. 
If it is diagonal, the algorithm returns the nodes in the tree that are on the path of the best decomposition and terminates.
Otherwise, the code enters three nested loops: one over all the columns, and two over indices that keep track of two different rows, where the innermost loop's index is always greater than the first loop over the rows. 
At every step of the execution of these loops the entries of the matrix indexed by the column counter and the two rows indices are selected and they are compared against two constants (indicated as as $0$ and \emph{threshold}).
Since the angle of rotation is calculated as a ratio between the two entries and the phase is calculated similarly with a ratio on the two components of each of these complex numbers, the thresholds guarantee that the coefficients will not go to infinity due to the usage of floating point numbers and possible division by a number close to zero. 
Moreover, the threshold should be chosen in relation to the experimental performance of the fundamental two-level couplings, such that the resulting parameter fluctuations are negligible compared to the inherent imperfections of the quantum operations.
Once the integrity of the entries is verified, the angle and phase of rotation are calculated as shown in the pseudo-code. 
These two parameters are then used to create the Givens-rotations of the decomposition as elementary blocks. 
These matrices represent the rotation on the Bloch sphere of two energy levels, but they are also the rotation of a subspace of the larger Hilbert space of the qudit.

\begin{algorithm}[t]
    \caption{Adaptive Decomposition}
    \label{alg:adaptive-composition}
    \begin{algorithmic}
        \Require Tree.root: $n$, $U$, Energy coupling Graph:$g$,\ Cost~Limit: $\mathit{cl}$ 
        \Ensure Decomposition, Best Cost 
        \If{$\mathit{checkDiagonal}( n.U )$}
            return 
        \EndIf
        \State $\mathit{father} = n$
        \For{$c$}
            \For{$r\geq c$}
                \For{$r_2 > r$}
                \If{$ U_{r,c}\neq 0 \text{ and } U_{r_2,c}> \mathit{threshold}$}
                    \State $\theta = 2\cdot\arctan[|U_{r_2,c}/U_{r,c}|]$
                    \State $\phi = -( \pi/2 + \arg[U_{r,c}] - \arg[U_{r_2,c}])$
                    \State $c',\: \pi\mathit{gates},\: g' = \mathit{Cost}(U, g, R)$
                    \State $ U' = R_{r,r_2}(\theta,\phi) \cdot U$
                    \If{$\mathit{father.cost}+c' < \mathit{cl}$} 
                        \State $\mathit{father.addNode}(\mathit{gates}, U', g', c', \mathit{cl} )$
                    \EndIf
                \EndIf
                \EndFor
            \EndFor
        \EndFor
        \For{\textit{child in father.children}}
        \State \textit{Adaptive(child)}
        \EndFor
   \end{algorithmic}
\end{algorithm}

The rotation between two levels $i$ and $j$ are expressed as 
\begin{align}
    R_{i,j}(\theta,\phi) = \exp\left(\frac{-i \theta}{2} \Big(\cos(\phi)\sigma_x^{i,j} + \sin(\phi)\sigma_y^{i,j}\Big)\right)\label{eq:SUPPlocalOps}
\end{align}
where $\sigma_{x,y}$ are the Pauli-$\{X,Y\}$ matrices, describing the two-level coupling, $\theta$ is the rotation angle, and $\phi$ is the phase of the rotation.

The cost of each rotation is calculated by a method that takes the necessary routing of the states on the graph of the qudit into account, as detailed in \autoref{sec:cost}.
Additionally, the reordering sequence that enables the rotation, and the modified graph are returned alongside the calculated cost.
If the total current costs of the decomposition is smaller than the cost limit, the rotations, the cost, the new unitary rotated, and the new graph are saved as new child of the current node in the tree.
The function recursively calls itself with each child of the current node as parameter. 
As the implemented method is in the category of backtracking algorithms, the worst case complexity is asymptotically exponential~\cite{btrack_complx}.
However, the experimental results show that a suitable choice of the cost function allows the algorithm to perform similarly to the \mbox{fixed QR decomposition}.

\subsection{Satisfying Energy Coupling Constraints}

The decomposition from the previous section led to a sequence of operations which are elementary on the respective technology, but not necessarily in-line with the level-to-state mapping, i.e.,~transitions between states might not be immediately possible.
This problem of satisfying the energy coupling constraints is addressed by inserting reordering gates into the operation sequence and introducing rules for correct patterned sequences.
\subsubsection{Routing}
The decomposition is driven by the cost for the possible logical rotations on the transition of two logical states.
Here, the energy coupling graph is used to track the position of each logical state. 
Consider a Givens-rotation on the logic states $i$ and $j$.
The returned routing sequence is a list of reordering gates that will bring the logical state $j$ adjacent to the logical state $i$ in the energy coupling graph. 
The reordering gates are constructed as Givens-rotations with default values $\theta = \pi,\phi = -\pi/2 $ and indices of rotation depending on the nodes on the shortest path from $i$ to $j$. The energy coupling graph is adapted accordingly. 
\autoref{alg:adaptive-composition} takes the energy coupling constraints into account.

\subsubsection{Graph rules}
Another challenge arises from the fact that the reordering pulses are not permutation matrices, but carry additional sign flips. This is a result of the mathematical form of the generators of the primitive physical rotations. In the following we outline a set of simple rules that have to be applied to the reordered gate sequence.
First, rotation matrices are defined to rotate from a lower level to a higher level. If during compilation we encounter a situation where this ordering is reversed, this leads to an inversion of the sign of the phase~$\phi$ for that operation.
Second, whenever a gate in the sequence is preceded by a reordering pulse whose higher level coincides with any of the levels of the gate in question, the sign of $\theta$ is flipped. 
The last rule leverages a particular feature of the graph, where each node is able to store a phase value that the physical level has accumulated during computation. 
The feature is useful for solving two challenges: the first one arises when compiling several unitaries in a row where the $Z$-rotations that are not expressed in the form of a gate can be just recorded in the graph's nodes, allowing for even shorter sequence and avoiding the problem of propagating these gates explicitly. 
The second challenge is unique to the adaptive algorithm where routing sequences are not uncomputed as in the QR~decomposition case. 
Consequently, the sign flips from the reordering pulses do not cancel and need to be tracked explicitly as an accumulation of $\pi$ phases due to the routing gates.
Hence, the third rule adds to the the phase value of a gate the phase stored on the higher energy node and subtracts the phase stored on the lower energy node.

\subsubsection{Ancilla states}
A further possibility enabled by the proposed approach is the inclusion of ancillary states in the graph. 
Use of ancillary states is known to enable improvements in gate complexity~\cite{Lanyon2008} and simplify the application of established quantum gates in qudit Hilbert spaces~\cite{ringbauer2021universal}.
After a state is declared as ancillary, the algorithm tracks them like regular states, but gains the option of exploiting them for routing purposes or caching.

\vspace{0.5em}
\begin{example}
    Consider the matrix in \autoref{eqn:hadamard-ancillas}, which is composed of two different portions. 
    The upper-left part is the Hadamard gate that we want to compile on the first 3 logic levels.
    The lower-right part is the matrix that will operate on two ancillary states.
    It is therefore possible to compile two separate operations in one unitary.
    
    \begin{align}\label{eqn:hadamard-ancillas}
        \setlength\arraycolsep{2pt}
        \begin{bmatrix}
        1 & 1 & 1 & 0 & 0  \\
        1 & e^{\frac{2\pi}{3}} & e^{\frac{-2\pi}{3}} & 0 & 0  \\
        1 & e^{\frac{-2\pi}{3}} & e^{\frac{2\pi}{3}}  & 0 & 0 \\
        0 & 0 & 0 & a_{11} & a_{12} \\
        0 & 0 & 0 & a_{21} & a_{22}  \\
        \end{bmatrix} 
    \end{align}
\end{example}

\vspace{0.4em}
\subsection{Phase Shift Propagation}
After decomposition and insertion of reordering gates from the previous sections, the phase shifts have to be taken into account.
The phase shifts are encoded in the diagonal matrix~$\Theta$ (see \autoref{eqn:decomposition}) and have to be decomposed into individual phase shifts on single levels to ensure correctness of the overall decomposition, albeit the phase shift are not physically realized.
These individual phase-shifts are formulated as diagonal matrices, with only the $i$-th entry on the diagonal set to a phase shift of $\phi$ of the logic state, and $1$ otherwise, with $i \in \{0,\ldots, d-1\}$.

\begin{example}
    Consider a three-level qudit system and a rotation of~$\phi$ on the second entry on the diagonal.
    The corresponding matrix is shown in \autoref{eqn:rotation-matrix}.
    \begin{align}\label{eqn:rotation-matrix}
        R_{Z,1}(\phi) = 
        \begin{bmatrix}
            1 & 0 & 0 \\
            0 & e^{i\phi} & 0 \\
            0 & 0 & 1 \\
        \end{bmatrix}
    \end{align}
\end{example}

These phase shifts can be introduced at zero cost in the sequence, since they can be propagated to the end of the circuit, where a phase shift is then negligible, as changes in phase immediately before a measurement do not change the outcome probabilities.
We refer to these rotations as \enquote{virtual $Z$-gates}\footnote{The term virtual $Z$-gate does not refer to the Pauli operation.} since they are not executed on the quantum computer.
The commutation of phase gates and Givens-rotations is exemplified in \autoref{eqn:commutation}.
\begin{align}\label{eqn:commutation}
    \setlength\arraycolsep{2pt}
    M
    \cdot
    R_{0,1}(\theta,\alpha) &= 
    R_{0,1}(\theta,\alpha-\phi+\gamma)
    \cdot
    M\\
    \text{with } M &= \begin{bmatrix}
        e^{i\phi} & 0 & 0 \\
        0 & e^{i\gamma} & 0 \\
        0 & 0 & e^{i\delta} \\
    \end{bmatrix} \notag
\end{align}
The equation is given for a rotation on the coupling $R_{0,1}$ but it can generalized to any coupling.
The application of the commutation can lead to a linear cost reduction, but is potentially limited once entangling gates are introduced into the circuit.

\subsection{Cost function}
\label{sec:cost}
The previous sections described the steps to decompose the target unitary, satisfy the energy coupling constraints, and propagate the phase shifts.
The decisions made in these steps are guided by a cost function, that is provided in this section.
While this cost function is inspired by the trapped-ion system recently presented in~\cite{ringbauer2021universal} and the numerical results derived from benchmarking the platform in the same reference, it is primarily determined by considerations that are found in a similar form also in other technologies.
More precisely:
\begin{itemize} 
    \item The cost of each rotation scales linearly with the rotation angle.
    \item Each two-level coupling is calibrated for a fixed rotation angle, typically $\nicefrac{\pi}{2}$. Angles that differ from $\nicefrac{\pi}{2}$ will incur additional cost due to non-ideal performance of physical devices. We assume this cost to increase linearly with the distance from the calibrated value.
    \item The system is initialized almost perfectly in a fixed initial state. Those couplings that do not involve the initial state will incur a larger cost (here we assume a factor, function of the distance of two logic states in the graph), since they require a more complicated multi-step calibration procedure. 
\end{itemize}
Taking into account these effects, we model the cost of the single rotation as
\begingroup
\begin{align}
    C_1 &= 10^{-4}\cdot  \mathit{d}(i,j) \cdot \big( 4\theta + \big|\operatorname{mod}(\theta +\frac{1}{4}, \frac{1}{2}) - \frac{1}{4}\big| \big) \label{eqn:cost}
\end{align}
\endgroup
for $\theta \geq \nicefrac{1}{4}$ (in units of $\pi$). 
Here the overall factor $10^{-4}$ comes from the fundamental cost of a $\pi$ rotation~\cite{ringbauer2021universal}.

Besides the bare cost of logical rotations, there is an additional cost associated with reordering the logical states on the coupling graph, since it is, in general, not fully connected. This is done with a sequence of $\pi$-rotations, with a cost again governed by \autoref{eqn:cost}. These two terms then constitute the full \emph{experimental cost} of a decomposition.

\section{Evaluation}
\label{sec:evaluation}

The proposed method of adaptive compilation utilizes the energy coupling graph to find more efficient decompositions.
To evaluate the method, we developed an implementation and compared it against the state-of-the-art fixed-sequence algorithm based on the QR factorization~\cite{ringbauer2021universal}.
The implementation is available under a free license via \href{https://github.com/cda-tum/qudit-compilation}{github.com/cda-tum/qudit-compilation}.
To account for different scenarios, we compile sets of random Clifford-unitaries~\cite{Clark2006}, which are central to quantum computing, for dimensions 3, 5 and 7 to three target architectures each. 
The architectures are different in the number of nodes, connectivity, and mapping of the logical states on the physical levels---resulting in nine different target architectures.
Diagonal unitaries are excluded from the sets, as they are decomposable in a sequence of virtual phase-gates and the cost of the decomposition in each algorithm is zero.
The evaluation was performed on a computer with an Intel i7-10750H and \SI{32}{\giga\byte} of memory, running Ubuntu 20.04 and Python~3.8.

The results are presented in \autoref{tab:results}.
The first column gives the dimensionality, followed by the number of considered unitaries in the second column.
The next three columns show the costs of the QR decomposition for the considered unitaries, listing the minimal, average, and maximal costs according to the cost function in \autoref{sec:cost}.
The same listing on costs are contained in the next three columns for the adaptive compilation.

\begin{table}[t]
    \centering
    \caption{Evaluation of costs for the QR and the proposed (adaptive) decomposition}
    \label{tab:results}
    \sisetup{group-separator={\,}, group-minimum-digits=4}
    \resizebox{\linewidth}{!}{\begin{tabular}{cS[table-format=4.0]S[table-format=3.0]S[table-format=3.2]S[table-format=3.0]@{\hskip 16pt}S[table-format=3.0]*2{S[table-format=3.2]}}
        \toprule
	    	              &           & \multicolumn{3}{c}{QR decomp.~\cite{ringbauer2021universal}} & \multicolumn{3}{c}{Proposed decomp.} \\\cmidrule(r{16pt}){3-5}\cmidrule{6-8}
		Dim  & {Unitaries}    & {min} & {avg}       & {max}              & {min} & {avg} & {max} \\ \midrule
	    3 & 333 & 8 & 25,47 & 140 & 4 & 21,69 & 36  \\ 
        ~ & 333 & 4 & 21,19 & 28 & 4 & 9,49 & 12 \\ 
        ~ & 333 & 12 & 35,89 & 44 & 4 & 14,93 & 20 \\\addlinespace
        
        5 & 2985 & 60 & 166,34 & 200 & 64 & 113,36 & 124,63  \\
        ~ & 2985 & 36 & 101,27 & 128 & 16 & 82,86 & 91,05  \\ 
        ~ & 2985 & 88 & 240,62 & 280 & 16 & 121,97 & 133,7 \\ \addlinespace
        
        7 & 6438 & 132 & 662,35 & 756 & 44 & 255,58 & 279,08  \\ 
        ~ & 6438 & 192 & 518,23 & 620 & 48 & 389,94 & 425,51  \\ 
        ~ &  6438	&  88  & 392,35	& 508 &	44 & 330,3	& 361,51\\ \bottomrule
    \end{tabular}}
    
    {\vspace{0.5em}\raggedright Cost is multiplied by $10^4$ to aid legibility.\par}
\end{table}

Comparing the costs in \autoref{tab:results} shows a clear advantage of the proposed adaptive algorithm.
In every case, the maximal cost of the considered set of unitaries is lower compared to that of the QR~decomposition---in some cases there is an improvement of up to a factor of 3.
The same conclusion can be drawn from the average costs as well, which is considerably lower in each set of considered unitaries.
Regarding the run-time, each decomposition of a unitary in this evaluation took less than \SI{1}{\second} for the QR decomposition and the adaptive approach, respectively. Hence, by spending the same computational efforts, we get significantly better compilation results.

Moreover, the proposed method can be configured to get even better results. 
In the evaluations summarized above, we set the cost~limit (see~\autoref{alg:adaptive-composition}) to 1.1 of the cost of the QR decomposition to have an acceptable run-time for the thousands of unitaries.
For a more realistic scenario, where a designer wants to realize an individual unitary, decreasing the cost limit will push the proposed algorithm to get even smaller costs---at the expense of longer run-time.
Furthermore, the improvement attainable by the proposed adaptive method of course depends on the unitary that is considered, but also on particularities of the underlying technology.
One important factor is how dense the connectivity of the energy levels is, where a denser (and more physically unrealistic) connectivity leads to less improvement on the cost by the adaptive algorithm. With the implementation of the proposed method being publicly available, every designer can accordingly use these features to get better results than the ones summarized in the extensive evaluations covering thousands of unitaries from above.

Overall, the results confirm that, using the proposed method, more efficient decompositions can be determined and, additionally, further option exists to trade-off worst-case costs and run-time---providing a solution that aims at fully exploiting the potential provided by the given technology.

\section{Conclusion}
\label{sec:conclusion}
In this work, we proposed an \emph{adaptive compilation} algorithm to attain efficient decompositions of arbitrary local unitaries for multi-level quantum systems. 
Compared to binary quantum systems, qudits feature a much richer state-space, allowing for different placements of logical information under consideration of practical constraints, such as the energy coupling constraints.
With the constraints and options represented in an \emph{energy coupling graph}, the proposed algorithm exploits the available potential by flexibly adapting the mapping between logical states and physical energy levels to enable a much more efficient compilation. 
Indeed, turning the argument around, the proposed algorithm provides a way to compare different placements of logical states by means of the average cost to decompose random Clifford unitaries. 
This flexibility of the proposed algorithm to adapt to realistic constraints in the possible couplings as well as the cost of different rotations enable significant improvements in gate costs.

\section*{Acknowledgements}
This project has received funding from the European Research Council (ERC) under the European Union’s Horizon 2020 research and innovation programme (grant agreement No. 101001318).
It is part of the Munich Quantum Valley, which is supported by the Bavarian state government with funds from the Hightech Agenda Bayern Plus and was partially supported by the BMK, BMDW, and the State of Upper Austria in the frame of the COMET program (managed by the FFG). This project has received funding from the European Union’s Horizon 2020 research and innovation programme under the Marie Sk{\l}odowska-Curie grant agreement No 840450.

\printbibliography
\end{document}